\documentclass[12pt]{article}
\usepackage{epsfig}
\newcommand\bea{\begin{eqnarray}}
\newcommand\eea{\end{eqnarray}}
\begin{document}
\thispagestyle{empty}
\bibliographystyle{unsrt}
\setlength{\baselineskip}{18pt}
\parindent 24pt
\vspace{50pt}

\begin{center}{
{\Large {{\bf Quantum decoherence and classical correlations  \\
of the harmonic oscillator in the Lindblad theory}} }
\vskip 1truecm
A. Isar ${^{1,2,(a)}}$ and W. Scheid ${^2}$  \\
$^1$ {\it Department of Theoretical Physics, Institute of
Physics and Nuclear Engineering,
Bucharest-Magurele, Romania }\\
$^2$ {\it Institut f\"{u}r Theoretische Physik der
Justus-Liebig-Universit\"{a}t,\\
Giessen, Germany} }
\end{center}

\vskip 1truecm
\begin{abstract}
In the framework of the Lindblad theory for open quantum
systems we determine the degree of quantum decoherence
and classical correlations of a harmonic oscillator
interacting with a thermal bath.  The transition from
quantum to classical behaviour of the considered system
is analyzed and it is shown that the classicality takes
place during a finite interval of time. We calculate also
the decoherence time and show that it has the same scale
as the time after which statistical fluctuations become
comparable with quantum fluctuations.
\end{abstract}

PACS numbers: 03.65.Yz, 05.30.-d

(a) e-mail address: isar@theory.nipne.ro

\section{Introduction}

The transition from quantum to classical physics and classicality of
quantum systems continue to be among the most interesting problems
in many fields of physics, for both conceptual and experimental
reasons \cite{gi96,pa01,zu03}. Two conditions are essential for the
classicality of a quantum system \cite{mo90,ha90}: a) quantum
decoherence (QD), that means the irreversible, uncontrollable and
persistent formation of a quantum correlation (entanglement) of the
system with its environment \cite{ali}, expressed by the damping of
the coherences present in the quantum state of the system, when the
off-diagonal elements of the density matrix of the system decay
below a certain level, so that this density matrix becomes
approximately diagonal and b) classical correlations (CC), expressed
by the fact that the Wigner function of the quantum system has a
peak which follows the classical equations of motion in phase space
with a good degree of approximation, that is the quantum state
becomes peaked along a classical trajectory. The necessity and
sufficiency of both QD and CC as conditions of classicality are
still a subject of debate. Both these conditions do not have an
universal character, so that they are not necessary for all physical
models. An important role in this discussion plays the temperature
of the environment and therefore it is worth to take into account
the differences between low and high temperature regimes. For
example, purely classical systems at very high temperatures are
described by a classical Fokker-Planck equation which does not
follow any trajectory in phase space (for very small kinetic energy,
compared to the thermal energy, when the probability distribution
becomes essentially independent of momentum), so that in this case
CC are not necessary. Likewise, one can have a classical behaviour
if the coherences are negligible, without having strong CC (for
example, in the case of a classical gas at finite temperature) and
the lack of strong correlations between the coordinate and its
canonical momentum does not necessarily mean that the system is
quantum. On the other hand, the condition of CC is not sufficient
for a system to become classical -- although the Wigner function can
show a sharp correlation in phase space, the quantum coherence never
vanishes for a closed system which has a unitary evolution.
Likewise, in the low temperature quantum regime one can observe
strong CC. For example, in the case of a purely damped quantum
harmonic oscillator (at zero temperature), the initial coherent
states remain coherent and perfectly follow classical trajectories
of a damped oscillator, but CC are not sufficient for classicality.

In the last two decades it has became more and more clear
that the classicality is an emergent property of open
quantum systems, since both main features of this process
-- QD and CC -- strongly depend on the interaction
between the system and its external environment
\cite{zu03,ze70,zu82,jo84,jo85,ca85,un89,hu92,pa93,zu91}.
A remarkable aspect of the current research helping in
understanding the nature of the quantum to classical
transition is that for the first time there have recently
been carried on experiments probing the boundary between
the quantum and the classical domains in a controlled way
\cite{br96,ch99,am98,kl98,tu00,ko01}.

The role of QD became relevant in many interesting
physical problems of field theory, atomic physics,
quantum optics, quantum information processing, quantum
gravity and cosmology, and condensed matter physics. We
mention here only a few of these problems
\cite{pa01,zu03}: to understand the way in which QD
favorizes the quantum to classical transition of density
fluctuations; to study systems of trapped and cold atoms
(or ions) which may offer the possibility of engineering
the environment, like trapped atoms inside cavities,
relation between decoherence and other cavity QED effects
(such as Casimir effect); on mesoscopic scale,
decoherence in the context of Bose-Einstein condensation.

In many cases one is interested in understanding the
specific causes of QD just because one wants to prevent
decoherence from damaging quantum states and to protect
the information stored in quantum states from the
degrading effect of the interaction with the environment.
Thus, decoherence is responsible for washing out the
quantum interference effects which are desirable to be
seen as signals in some experiments. QD has a negative
influence on many areas relying upon quantum coherence
effects, such as quantum computation and quantum control
of atomic and molecular processes. In the physics of
information and computation, decoherence is an obvious
major problem in the implementation of
information-processing hardware that takes advantage of
the superposition principle \cite{ni00}.

In most of literature, QD has been studied for a system
coupled to an environment or thermal bath with many
degrees of freedom. The main purpose of this paper is to
study QD and CC for a harmonic oscillator interacting
with an environment in the framework of the Lindblad
theory for open quantum systems. More concretely we
determine the degree of QD and CC and the possibility of
simultaneous realization of QD and CC for a system
consisting of a harmonic oscillator in a thermal bath.
For that purpose, we first find the evolution of the
density matrix and of the Wigner function of the
considered system and then we apply the criterion of QD
and CC. We consider different regimes of the temperature
of environment. It is found that the system manifests a
QD which increases with time and temperature, whereas CC
are less and less strong with increasing time and
temperature.

The organizing of the paper is as follows. In Sec. 2 we
review the Lindblad master equation for the damped
harmonic oscillator and in Sec. 3 we derive the master
equation in coordinate representation and the
corresponding Fokker-Planck equation in the Wigner
representation and determine the density matrix and
Wigner function of the considered system. Then in Sec. 4
we investigate QD and CC and analyze them quantitatively.
In Sec. 5 we calculate the decoherence time of the system
and discuss the transition from quantum mechanics to
classical statistical mechanics. A summary and concluding
remarks are given in Sec. 6.

\section{Lindblad master equation for the harmonic
oscillator}

Here we review the Lindblad's axiomatic formalism based
on quantum dynamical semigroups. The irreversible time
evolution of an open system is described by the following
general quantum Markovian master equation for the density
operator $\rho(t)$ \cite{d,l1,s}: \bea{d \rho(t)\over
dt}=-{i\over\hbar}[ H, \rho(t)]+{1\over 2\hbar}
\sum_{j}([  V_{j} \rho(t), V_{j}^\dagger ]+[ V_{j},
\rho(t) V_{j}^\dagger ]).\label{lineq}\eea $H$ is the
Hamiltonian of the system and $V_{j},$ $ V_{j}^\dagger $
are operators on the Hilbert space of $H$, which model
the environment. In order to obtain, for the damped
quantum harmonic oscillator, equations of motion as close
as possible to the classical ones, the two possible
operators $V_{1}$ and $ V_{2}$ are taken as linear
polynomials in coordinate $q$ and momentum $p$
\cite{l2,ss,rev} and the harmonic oscillator Hamiltonian
$H$ is chosen of the general quadratic form \bea
H=H_{0}+{\mu\over 2}(qp+pq), ~~~  H_{0}={1\over
2m}p^2+{m\omega^2\over 2}  q^2. \label{ham} \eea With
these choices the master equation (\ref{lineq}) takes the
following form \cite{ss,rev}: \bea {d \rho \over
dt}=-{i\over \hbar}[ H_{0}, \rho]- {i\over
2\hbar}(\lambda +\mu) [  q, \rho p+ p \rho]+{i\over
2\hbar}(\lambda -\mu)[  p,
\rho   q+  q \rho]  \nonumber\\
-{D_{pp}\over {\hbar}^2}[ q,[  q, \rho]]-{D_{qq}\over
{\hbar}^2} [ p,[  p, \rho]]+{D_{pq}\over {\hbar}^2}([
q,[  p, \rho]]+ [ p,[ q, \rho]]). ~~~~\label{mast}   \eea
The quantum diffusion coefficients $D_{pp},D_{qq},$
$D_{pq}$ and the dissipation constant $\lambda$ satisfy
the following fundamental constraints \cite{ss,rev}: $
D_{pp}>0, D_{qq}>0$ and \bea D_{pp}D_{qq}-D_{pq}^2\ge
{{\lambda}^2{\hbar}^2\over 4}. \label{ineq} \eea In the
particular case when the asymptotic state is a Gibbs
state $ \rho_G(\infty)=e^{-{  H_0\over kT}}/ {\rm
Tr}e^{-{  H_0\over kT}}, $ these coefficients become
\cite{ss,rev} \bea D_{pp}={\lambda+\mu\over 2}\hbar
m\omega\coth{\hbar\omega\over 2kT},
~~D_{qq}={\lambda-\mu\over 2}{\hbar\over
m\omega}\coth{\hbar\omega\over 2kT}, ~~D_{pq}=0,
\label{coegib} \eea where $T$ is the temperature of the
thermal bath. In this case, the fundamental constraints
are satisfied only if $\lambda>\mu$ and \bea
(\lambda^2-\mu^2)\coth^2{\hbar\omega\over 2kT}
\ge\lambda^2.\label{cons}\eea

From the master equation (\ref{mast}) we obtain the
following equations of motion for the expectation values
of coordinate and momentum \cite{ss,rev}:
\bea{d\sigma_{q}(t)\over
dt}=-(\lambda-\mu)\sigma_{q}(t)+{1\over m}\sigma_{p} (t),
\label{eqmo1}\eea \bea{d\sigma_{p}(t)\over
dt}=-m\omega^2\sigma_{q}(t)-(\lambda+\mu)\sigma_{p}(t).
\label{eqmo2}\eea In the underdamped case $(\omega>\mu)$
considered in this paper, with the notation
$\Omega^2\equiv\omega^2-\mu^2$, we obtain \cite{ss,rev}:
\bea\sigma_q(t)=e^{-\lambda t}((\cos\Omega
t+{\mu\over\Omega}\sin\Omega t) \sigma_q(0)+{1\over
m\Omega}\sin\Omega t\sigma_p(0)), \label{sol1}\eea
\bea\sigma_p(t)=e^{-\lambda
t}(-{m\omega^2\over\Omega}\sin\Omega t\sigma_q(0)+
(\cos\Omega t-{\mu\over\Omega}\sin\Omega t)\sigma_p(0))
\label{sol2}\eea and
$\sigma_q(\infty)=\sigma_p(\infty)=0.$

Lindblad has proven \cite{l2} that in the Markovian
regime the harmonic oscillator master equation which
satisfies the complete positivity condition cannot
satisfy simultaneously the translational invariance and
the detailed balance (which assures an asymptotic
approach to the canonical thermal equilibrium state). The
necessary and sufficient condition for translational
invariance is $\lambda=\mu$ \cite{l2,ss,rev}. In this
case the equations of motion (\ref{eqmo1}) and
(\ref{eqmo2}) are exactly the same as the classical ones.
If $\lambda\neq \mu,$ then we violate translational
invariance, but we keep the canonical equilibrium state.

The relation (\ref{ineq}) is a necessary condition for
the generalized uncertainty inequality \bea
\sigma_{qq}(t)\sigma_{pp}(t)-\sigma_{pq}^2(t)\ge{\hbar^2
\over 4}\label{genun1}\eea to be fulfilled, where
$\sigma_{qq}$ and $\sigma_{pp}$ denote the dispersion
(variance) of the coordinate and momentum, respectively,
and $\sigma_{pq}$ denotes the correlation (covariance) of
the coordinate and momentum. The equality in relation
(\ref{genun1}) is realized for a special class of pure
states, called correlated coherent states \cite{dodkur}
or squeezed coherent states.

The asymptotic values $\sigma_{qq}(\infty),
\sigma_{pp}(\infty),\sigma_{pq}(\infty)$ do not depend on the
initial values $\sigma_{qq}(0),\sigma_{pp}(0),\sigma_{pq}(0)$ and in
the case of a thermal bath with coefficients (\ref{coegib}), they
reduce to \cite{ss,rev} \bea \sigma_{qq}(\infty)={\hbar\over
2m\omega}\coth{\hbar\omega\over 2kT}, ~~\sigma_{pp}(\infty)={\hbar
m\omega\over 2}\coth{\hbar\omega\over 2kT}, ~~\sigma_{pq}(\infty)=0.
\label{varinf}\eea

In the following, we consider a general temperature $T,$ but we
should stress that the Lindblad theory is obtained in the Markov
approximation, which holds for high temperatures of the environment.
At the same time, the semigroup dynamics of the density operator
which must hold for a quantum Markovian process is valid only for
the weak-coupling regime, with the damping $\lambda$ obeying the
inequality $\lambda\ll\omega.$

\section{Density matrix and Wigner distribution function}

We consider a harmonic oscillator with an initial Gaussian wave
function \bea \Psi(q)=({1\over 2\pi\sigma_{qq}(0)})^{1\over
4}\exp[-{1\over 4\sigma_{qq}(0)}
(1-{2i\over\hbar}\sigma_{pq}(0))(q-\sigma_q(0))^2+{i\over
\hbar}\sigma_p(0)q], \label{ccs}\eea where $\sigma_{qq}(0)$ is the
initial spread, $\sigma_{pq}(0)$ the initial covariance, and
$\sigma_q(0)$ and $\sigma_p(0)$ are the initial averaged position
and momentum of the wave packet. The initial state (\ref{ccs})
represents a correlated coherent state \cite{dodkur} with the
variances and covariance of coordinate and momentum \bea
\sigma_{qq}(0)={\hbar\delta\over 2m\omega},~~ \sigma_{pp}(0)={\hbar
m\omega\over 2\delta(1-r^2)},~~ \sigma_{pq}(0)={\hbar r\over
2\sqrt{1-r^2}}. \label{inw}\eea Here, $\delta$ is the squeezing
parameter which measures the spread in the initial Gaussian packet
and $r,$ with $|r|<1$ is the correlation coefficient at time $t=0.$
The initial values (\ref{inw}) correspond to a minimum uncertainty
state, since they fulfil the generalized uncertainty relation \bea
\sigma_{qq}(0)\sigma_{pp}(0)-\sigma_{pq}^2(0) ={\hbar^2\over
4}.\label{gen0}\eea For $\delta=1$ and $r=0$ the correlated coherent
state becomes a Glauber coherent state. For a given temperature $T$
of the bath and for any parameters $\delta$ and $r$ the inequality
(\ref{cons}) alone determines the range of values of the parameters
$\lambda$ and $\mu$ \cite{unc}.

From Eq. (\ref{mast}) we derive the evolution equation in
coordinate representation: \bea {\partial\rho\over\partial
t}={i\hbar\over 2m}({\partial^2\over\partial q^2}-
{\partial^2\over\partial q'^2})\rho-{im\omega^2\over
2\hbar}(q^2-q'^2)\rho\nonumber\\
-{1\over 2}(\lambda+\mu)(q-q')({\partial\over\partial
q}-{\partial\over\partial q'})\rho+{1\over
2}(\lambda-\mu)[(q+q')({\partial\over\partial
q}+{\partial\over\partial
q'})+2]\rho  \nonumber\\
-{D_{pp}\over\hbar^2}(q-q')^2\rho+D_{qq}({\partial\over\partial
q}+{\partial\over \partial q'})^2\rho -{2iD_{pq}\hbar}(q-q')(
{\partial\over\partial q}+{\partial\over\partial
q'})\rho\label{cooreq}\eea and in Refs. \cite{i2,wig,vlas} we
transformed the master equation (\ref{mast}) for the density
operator into the following Fokker-Planck-type equation satisfied by
the Wigner distribution function $W(q,p,t):$ \bea   {\partial
W\over\partial t}= -{p\over m}{\partial W\over\partial q} +m\omega^2
q{\partial W\over\partial p} +(\lambda+\mu){\partial\over\partial
p}(pW)
+(\lambda-\mu){\partial\over\partial q}(qW) \nonumber \\
+D_{pp}{\partial^2 W\over\partial p^2} +D_{qq}{\partial^2
W\over\partial q^2} +2D_{pq}{\partial^2 W\over\partial p\partial
q}.~~~~~~~~~~~~~~~~~~~ \label{wigeq}\eea The first two terms on the
right-hand side of both these equations generate a purely unitary
evolution. They give the usual Liouvillian evolution. The third and
forth terms are the dissipative terms and have a damping effect
(exchange of energy with environment). The last three are noise
(diffusive) terms and produce fluctuation effects in the evolution
of the system. $D_{pp}$ promotes diffusion in momentum and generates
decoherence in coordinate $q$: it reduces the off-diagonal terms,
responsible for correlations between spatially separated pieces of
the wave packet. Similarly $D_{qq}$ promotes diffusion in coordinate
and generates decoherence in momentum $p.$ The $D_{pq}$ term is the
so-called "anomalous diffusion" term. It promotes diffusion in the
variable $qp+pq,$ just like both the other diffusion terms, but it
does not generate decoherence.

In the high temperature limit, quantum Fokker-Planck equation
(\ref{wigeq}) with coefficients (\ref{coegib}) becomes classical
Kramers equation ($D_{pp}\to 2m\lambda kT$ for $\lambda=\mu$)
\cite{vlas}.

The density matrix solution of Eq. (\ref{cooreq}) has the general
form of Gaussian density matrices \bea <q|\rho(t)|q'>=({1\over
2\pi\sigma_{qq}(t)})^{1\over 2} \exp[-{1\over
2\sigma_{qq}(t)}({q+q'\over
2}-\sigma_q(t))^2\nonumber\\
-{\sigma(t)\over 2\hbar^2\sigma_{qq}(t)}(q-q')^2
+{i\sigma_{pq}(t)\over \hbar\sigma_{qq}(t)}({q+q'\over
2}-\sigma_q(t))(q-q')+{i\over
\hbar}\sigma_p(t)(q-q')],\label{densol} \eea where
\bea\sigma(t)\equiv\sigma_{qq}(t)\sigma_{pp}(t)-\sigma_{pq}^2(t)
\label{det}\eea is the determinant of the dispersion (correlation)
matrix \bea \left(\matrix{\sigma_{qq}(t)&\sigma_{pq}(t)\cr
\sigma_{pq}(t)&\sigma_{pp}(t)\cr}\right)\label{sigma}\eea and
represents also the Schr\"odinger generalized uncertainty function
\cite{unc}.

For an initial Gaussian Wigner function (corresponding to a
correlated coherent state (\ref{ccs})) the solution of Eq.
(\ref{wigeq}) is \bea W(q,p,t)={1\over 2\pi\sqrt{\sigma(t)}}
\exp\{-{1\over 2\sigma(t)}[\sigma_{pp}(t)(q-\sigma_q(t))^2+
\sigma_{qq}(t)(p-\sigma_p(t))^2\nonumber\\
-2\sigma_{pq}(t)(q-\sigma_q(t))(p-\sigma_p(t))]
\}.\label{wig} \eea

In the case of a thermal bath we obtain the following steady state
solution for $t\to\infty$ (we denote
$\epsilon\equiv{\hbar\omega\over 2kT}$): \bea
<q|\rho(\infty)|q'>=({m\omega\over \pi\hbar\coth\epsilon})^{1\over
2}\exp\{-{m\omega\over 4\hbar}[{(q+q')^2\over\coth\epsilon}+
(q-q')^2\coth\epsilon]\}.\label{dinf}\eea In the long time limit we
have also \bea W_{\infty}(q,p)={1\over \pi\hbar\coth
\epsilon}\exp\{-{1\over \hbar\coth\epsilon}[m\omega q^2+{p^2\over
m\omega}] \}.\label{wiginf} \eea Stationary solutions to the
evolution equations obtained in the long time limit are possible as
a result of a balance between the wave packet spreading induced by
the Hamiltonian and the localizing effect of the Lindblad operators.

\section{Quantum decoherence and classical correlations}

As we stated in the Introduction, one considers that two conditions
have to be satisfied in order that a system could be considered as
classical. The {\it first} condition requires that the system should
be in one of relatively permanent states (states that are least
affected by the interaction of the system with the environment,
called by Zurek "preferred states" in the environment induced
superselection description \cite{pa01,zu03}) and the interference
between different states should be negligible. This implies the
destruction of off-diagonal elements representing coherences between
quantum states in the density matrix, which is the QD phenomenon. An
isolated system has an unitary evolution and the coherence of the
state is not lost -- pure states evolve in time only to pure states.
The loss of coherence can be achieved by introducing an interaction
between the system and environment: an initial pure state with a
density matrix which contains nonzero off-diagonal terms can
non-unitarily evolve into a final mixed state with a diagonal
density matrix during the interaction with the environment, like in
classical statistical mechanics.

The {\it second} condition requires that the system should have,
with a good approximation, an evolution according to classical laws.
This implies that the Wigner distribution function has a peak along
a classical trajectory, that means there exist CC between the
canonical variables of coordinate and momentum. Of course, the
correlation between the canonical variables, necessary to obtain a
classical limit, should not violate Heisenberg uncertainty
principle, i.e. the position and momentum should take reasonably
sharp values, to a degree in concordance with the uncertainty
principle. This is possible, because the density matrix does not
diagonalize exactly in position, but with a non-zero width, i.e. it
is strongly peaked about $q=q'$ and very small for $q$ far from
$q'.$

Using new variables $\Sigma=(q+q')/2$ and $\Delta=q-q',$
the density matrix (\ref{densol}) can be rewritten as
\bea \rho(\Sigma,\Delta,t)=\sqrt{\alpha\over
\pi}\exp[-\alpha\Sigma^2
-\gamma\Delta^2+i\beta\Sigma\Delta+2\alpha\sigma_q(t)\Sigma
+i({\sigma_p(t)\over\hbar}-
\beta\sigma_q(t))\Delta-\alpha\sigma_q^2(t)],\label{ccd3}\eea
with the abbreviations \bea \alpha={1\over
2\sigma_{qq}(t)},~~\gamma={\sigma(t)\over 2\hbar^2
\sigma_{qq}(t)},~~
\beta={\sigma_{pq}(t)\over\hbar\sigma_{qq}(t)}\label{ccd4}\eea
and the Wigner transform of the density matrix
(\ref{ccd3}) is \bea W(q,p,t)={1\over
2\pi\hbar}\sqrt{\alpha\over\gamma}\exp\{-{[\hbar\beta
(q-\sigma_q(t))-(p-\sigma_p(t))]^2\over
4\hbar^2\gamma}-\alpha (q-\sigma_q(t))^2\}.\label{wigc}
\eea

a) {\it Degree of quantum decoherence (QD}

The representation-independent measure of the degree of QD
\cite{mo90} is given by the ratio of the dispersion
$1/\sqrt{2\gamma}$ of the off-diagonal element $\rho(0,\Delta,t)$ to
the dispersion $\sqrt{2/\alpha}$ of the diagonal element
$\rho(\Sigma,0,t):$ \bea \delta_{QD}={1\over 2}\sqrt{\alpha\over
\gamma},\label{qdec}\eea which in our case gives \bea
\delta_{QD}(t)={\hbar\over 2\sqrt{\sigma(t)}}.\eea

The finite temperature Schr\"odinger generalized uncertainty
function (\ref{det}), calculated in Ref. \cite{unc}, has the
expression \bea\sigma(t)={\hbar^2\over 4}\{e^{-4\lambda
t}[1-(\delta+{1\over\delta(1-r^2)})\coth\epsilon+\coth^2\epsilon]
\nonumber\\
+e^{-2\lambda
t}\coth\epsilon[(\delta+{1\over\delta(1-r^2)}
-2\coth\epsilon){\omega^2-\mu^2\cos(2\Omega
t)\over\Omega^2}\nonumber \\
+(\delta-{1\over\delta(1-r^2)}){\mu \sin(2\Omega
t)\over\Omega}+{2r\mu\omega(1-\cos(2\Omega
t))\over\Omega^2\sqrt{1-r^2}}]+\coth^2\epsilon\}.\label{sunc}\eea In
the limit of long times Eq. (\ref{sunc}) yields \bea
\sigma(\infty)={\hbar^2\over 4}\coth^2\epsilon,\eea so that we
obtain \bea \delta_{QD}(\infty)=\tanh{\hbar\omega\over
2kT},\label{fqd}\eea which for high $T$ becomes
\bea\delta_{QD}(\infty)={\hbar\omega\over 2kT}.\eea

We see that $\delta_{QD}$ decreases, and therefore QD increases,
with temperature, i.e. the density matrix becomes more and more
diagonal at higher $T$ and the contributions of the off-diagonal
elements get smaller and smaller. At the same time the degree of
purity decreases and the degree of mixedness increases with $T.$
$\delta_{QD}<1$ for $T\neq 0,$ while for $T=0$ the asymptotic
(final) state is pure and $\delta_{QD}$ reaches its initial maximum
value 1. A pure state undergoing unitary evolution is highly
coherent: it does not lose its coherence, i.e. off-diagonal
coherences never vanish. $\delta_{QD}= 0$ when the quantum coherence
is completely lost. So, when $\delta_{QD}= 1$ there is no QD and
only if $\delta_{QD}<1,$ there is a significant degree of QD, when
the magnitude of the elements of the density matrix in the position
basis are peaked preferentially along the diagonal $q=q'.$ When
$\delta_{QD}\ll 1,$ we have a strong QD.

b) {\it Degree of classical correlations (CC)}

In defining the degree of CC, the form of the Wigner function is
essential, but not its position around $\sigma_q(t)$ and
$\sigma_p(t).$ Consequently, for simplicity we consider zero values
for the initial expectations values of the coordinate and momentum
and the expression (\ref{wigc}) of the Wigner function becomes \bea
W(q,p,t)={1\over
2\pi\hbar}\sqrt{\alpha\over\gamma}\exp[-{(\hbar\beta q-p)^2\over
4\hbar^2\gamma}-\alpha q^2].\label{wigcs} \eea A ridge of the Wigner
function (\ref{wigcs}) in phase space is at $p=\hbar\beta q,$
showing the correlation between $q$ and $p.$ As a measure of the
degree of CC we take the relative sharpness of this peak in the
phase space determined from the dispersion $\hbar\sqrt{2\gamma}$ in
$p$ in Eq. (\ref{wigcs}) and the magnitude of the average of $p$
($p_0=\hbar\beta q $) \cite{mo90}: \bea
\delta_{CC}={2\sqrt{\alpha\gamma}\over|\beta|},\label{cor}\eea where
we identified $q$ as the dispersion $1/\sqrt{2\alpha}$ of $q.$
$\delta_{CC}$ is a good measure of the "squeezing" of the Wigner
function in phase space \cite{mo90}: in the state (\ref{wigcs}),
more "squeezed" is the Wigner function, more strongly established
are CC. In the coordinates $\hbar\beta q-p$ and $\hbar\beta q$
(these quantities have the same dimension), $2\hbar\sqrt{\gamma}$
and $\hbar|\beta|/\sqrt{\alpha}$ are the lengths of the shorter and
longer semi-axes of the 1$\sigma$ contour in phase space and their
ratio gives $\delta_{CC}.$ Similarly, in coordinates $\hbar\beta
q-p$ and $q,$  $2\hbar\sqrt{\gamma}$ and $1/\sqrt{\alpha}$ are the
lengths of the shorter and longer semi-axes of the 1$\sigma$ contour
and their product gives the area of the $1\sigma$ ellipse. We see
from Eq. (\ref{qdec}) that $\delta_{QD}$ is inversely proportional
to this area. Besides this geometric interpretation, $\delta_{QD}$
is also connected with the linear entropy \cite{for,pur}.

For our case, we obtain \bea \delta_{CC}(t)={\sqrt{\sigma(t)}\over
|\sigma_{pq}(t)|},\eea where $\sigma(t)$ is given by Eq.
(\ref{sunc}) and $\sigma_{pq}(t)$ can be calculated using formulas
given in Refs. \cite{ss,rev}: \bea\sigma_{pq}(t)={\hbar\over
4\Omega^2}e^{-2\lambda
t}\{[\mu\omega(2\coth\epsilon-\delta-{1\over\delta(1-r^2)})
-{2\omega^2r\over\sqrt{1-r^2}}]\cos(2\Omega t)\nonumber \\
+\omega\Omega(\delta-{1\over\delta(1-r^2)})\sin(2\Omega
t)+\mu\omega(\delta+{1\over\delta(1-r^2)}-2\coth\epsilon)+{2\mu^2r
\over\sqrt{1-r^2}}\}.\label{pqvar}\eea When $\delta_{CC}$
is of order of unity, we have a significant degree of
classical correlations. The condition of strong CC is
$\delta_{CC}\ll 1,$ which assures a very sharp peak in
phase space. Since $\sigma_{pq}(\infty)=0,$ in the case
of an asymptotic Gibbs state, we get
$\delta_{CC}(\infty)\to\infty,$ so that our expression
shows no CC at $t\to\infty.$

c) {\it Discussion with Gaussian density matrix and Wigner function}

We have seen that if the initial wave function is
Gaussian, then the density matrix (\ref{densol}) and the
Wigner function (\ref{wig}) remain Gaussian for all times
(with time-dependent parameters which determine their
amplitude and spread) and centered along the trajectory
given by Eqs. (\ref{sol1}) and (\ref{sol2}), which are
the solutions $\sigma_q(t)$ and $\sigma_p(t)$ of the
dissipative equations of motion (\ref{eqmo1}) and
(\ref{eqmo2}). This trajectory is exactly classical for
$\lambda=\mu$ and only approximately classical for not
large $\lambda-\mu.$ In Fig. 1 there are represented the
trajectory in phase space and two examples of the
1$\sigma$ contour of the initial Wigner function,
corresponding to an initial coherent state ($\delta=1$)
and a squeezed state ($\delta=4$). In general, the
$1\sigma$ contour is defined by the ellipse \bea {1\over
2\sigma(t)}[\sigma_{pp}(t)(q-\sigma_q(t))^2+
\sigma_{qq}(t)(p-\sigma_p(t))^2
-2\sigma_{pq}(t)(q-\sigma_q(t))(p-\sigma_p(t))]=1.\eea In
Fig. 1 the center of the ellipses is the point given by
the initial expectation values $\sigma_q(0)$ and
$\sigma_q(0).$

\begin{figure}
\label{Fig. 1} \centerline{\epsfig{file=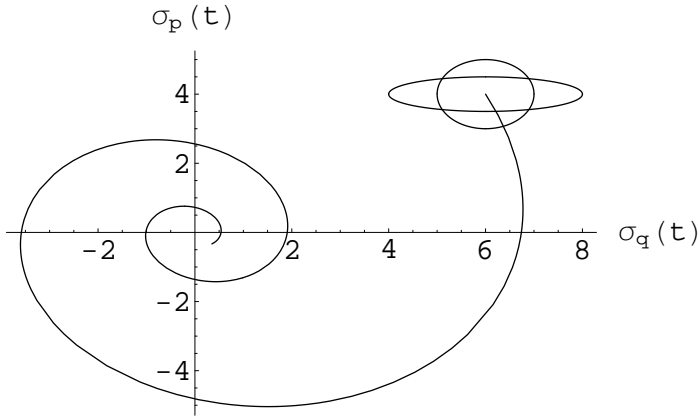,
width=0.6\textwidth}} \caption{Trajectory in phase space
given by the expectation values of coordinate
$\sigma_q(t)$ and momentum $\sigma_p(t)$ for time $t\in
[0,14]$ with initial coordinate $\sigma_q(0)=6$ and
momentum $\sigma_p(0)=4$ and 1$\sigma$ contours of the
Wigner function, corresponding to an initial coherent
state ($\delta=1$) and a squeezed state ($\delta=4$), for
$\lambda=0.2$ and $\mu=0.1.$ In all figures we use the
system of units $m=\omega=\hbar=1.$}
\end{figure}

To illustrate the dependence on the temperature and time of the
degree of QD and the degree of CC, we represent them in Fig. 2. We
can see that the degree of QD has an evolution which shows that in
general QD increases with time and temperature. We can also see that
the degree of CC has a more complicated evolution, but the general
tendency is that CC are less and less strong with increasing time
and temperature. $\delta_{QD}<1$ for non-zero temperature and
$\delta_{CC}$ is of the order of unity for a long enough interval of
time, so that we can say that the considered system interacting with
the thermal bath manifests both QD and CC and a true quantum to
classical transition takes place. Dissipation promotes quantum
coherences, whereas fluctuation (diffusion) reduces coherences and
promotes QD. The balance of dissipation and fluctuation determines
the final equilibrium value of $\delta_{QD}.$ The quantum system
starts as a pure state, with a Wigner function well localized in
phase space (Gaussian form). This state evolves approximately
following the classical trajectory (Liouville flow) in phase space
and becomes a quantum mixed state during the irreversible process of
QD.

\begin{figure}
\label{Fig. 2} \centerline{\epsfig{file=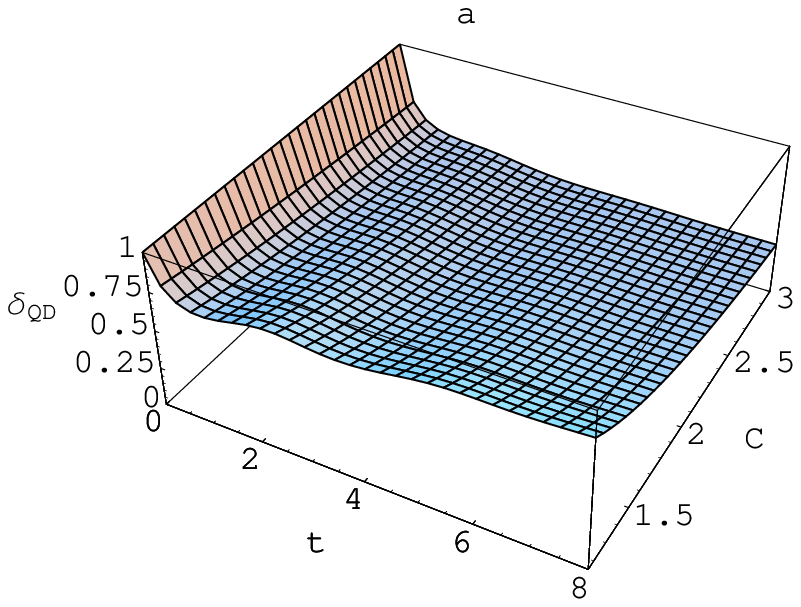,
width=0.6\textwidth}} \centerline{\epsfig{file=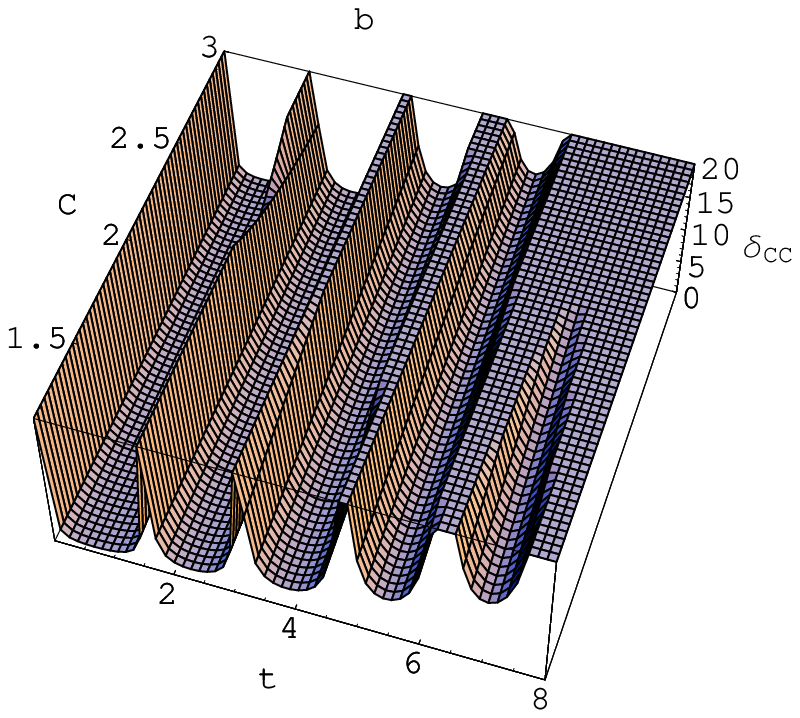,
width=0.6\textwidth}}\caption{a: Degree of quantum
decoherence $\delta_{QD}$ and b: degree of classical
correlations $\delta_{CC}$ as functions on temperature
$T$ (through $C\equiv\coth{\hbar\omega\over 2kT}$) and
time $t$ for $\lambda=0.2,\mu=0.1,\delta=4,r=0.$}
\end{figure}

The squeezing and the correlation of the initial state
play also a role in the degree of QD and CC. When the
squeezing parameter $\delta$ is increasing, both
$\delta_{QD}$ and $\delta_{CC}$ are decreasing, therefore
the squeezing favorizes both QD and CC. Likewise, the
increasing of the correlation coefficient $r$ leads to
the decreasing of the degree of QD, so that when we have
a larger initial correlation, then QD is stronger. At the
same time, the variation of the correlation coefficient
has a small influence on the general pulsatory behaviour
of the degree of CC. We also remark from Eq. (\ref{fqd})
that the asymptotic value of the degree of QD does not
depend on the initial squeezing and correlation, it
depends on temperature only.

In Figs. 3 and 4 we represent the density matrix in
coordinate representation (\ref{densol}) and the Wigner
function (\ref{wig}) at the initial and final moments of
time. The values of the density matrix along the diagonal
$q=q'$ represent the probability of finding the system in
this position, while the off-diagonal values represent
the correlations in the density matrix between the points
$q$ and $q'.$ The asymptotic Wigner distribution has an
axial symmetry, reflecting quantum equipartition. For
simplicity, in Figs. 3 and 4 we consider zero values for
the initial expectations values of coordinate and
momentum, so that both the density matrix and Wigner
function are centered in origin. Of course, as we stated
earlier, for non-zero initial expectations values of
coordinate and momentum, the density matrix
(\ref{densol}) and the Wigner function (\ref{wig}) are
centered along the trajectory given by Eqs. (\ref{sol1})
and (\ref{sol2}), like in Fig. 1.

\begin{figure}
\label{Fig. 3} \centerline{\epsfig{file=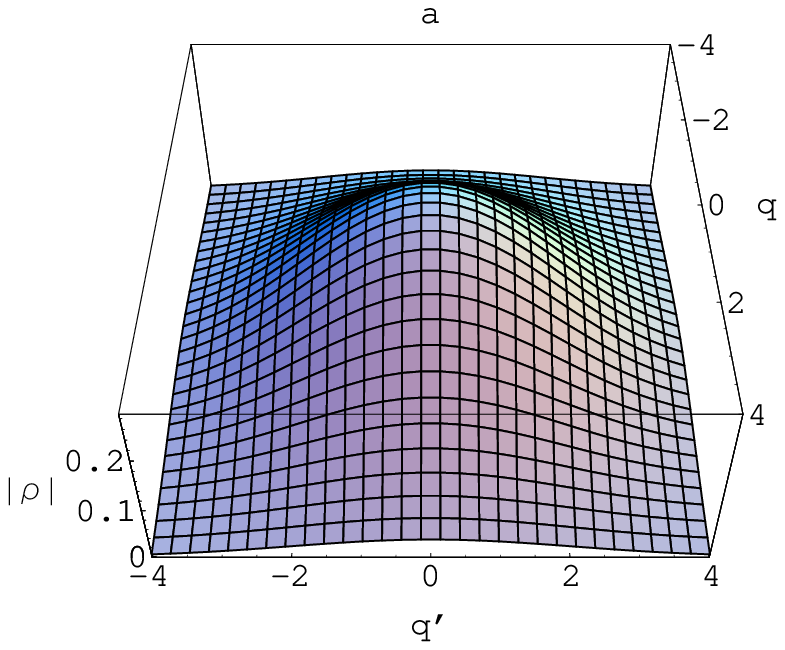,
width=0.55\textwidth}}
\centerline{\epsfig{file=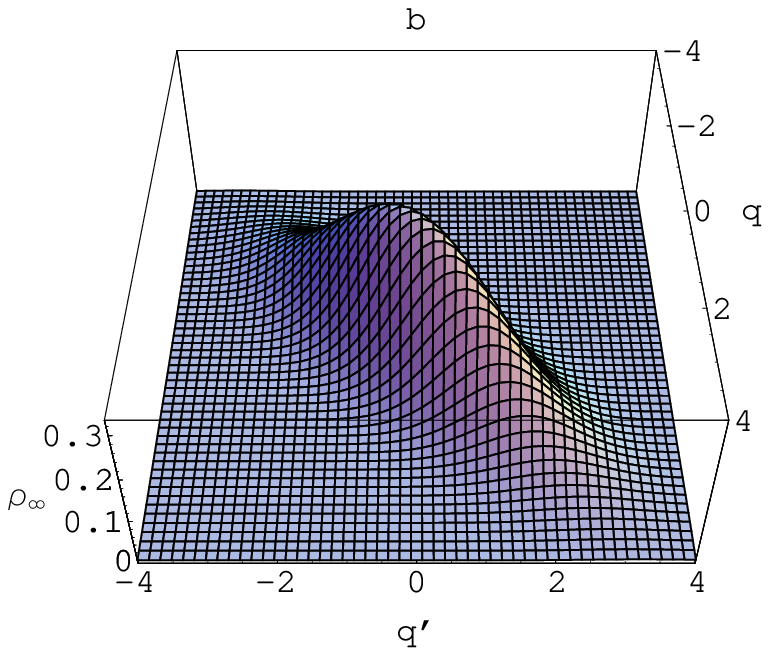,
width=0.55\textwidth}}\centerline{\epsfig{file=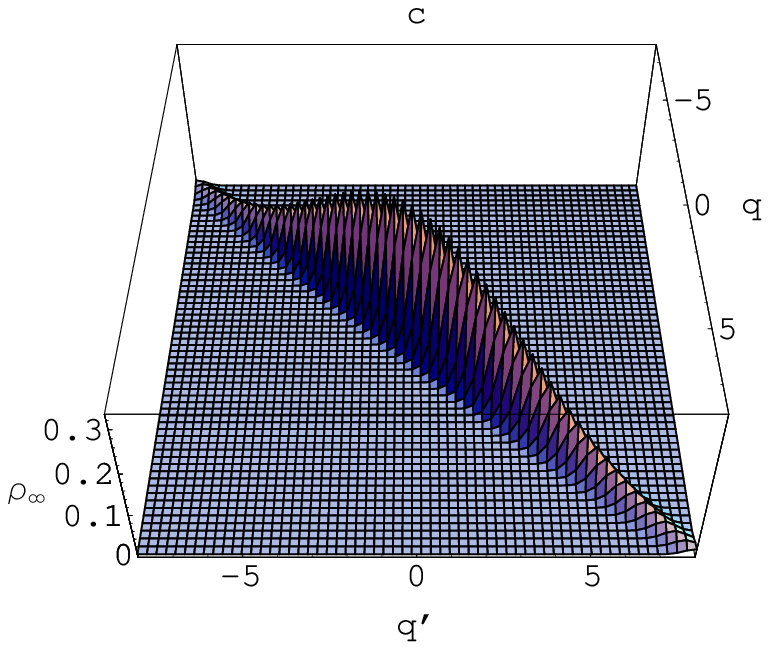,
width=0.55\textwidth}}\caption{Density matrix $\rho$ in
coordinate representation for
$\lambda=0.2,\mu=0.1,\delta=4,r=0;$ a: $|\rho|$ at the
initial time $t=0;$ b: $\rho_{\infty}$ at time
$t\to\infty$ for $C=3;$ c: $\rho_{\infty}$ for $C=20.$}
\end{figure}

\begin{figure}
\label{Fig. 4} \centerline{\epsfig{file=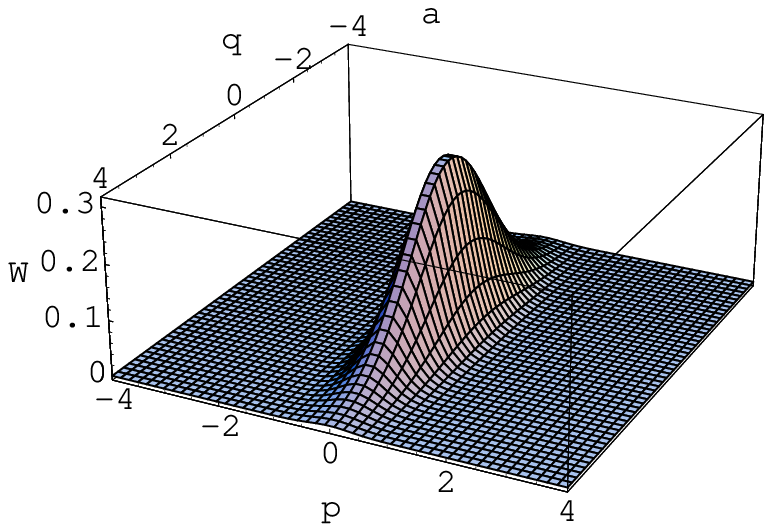,
width=0.6\textwidth}} \centerline{\epsfig{file=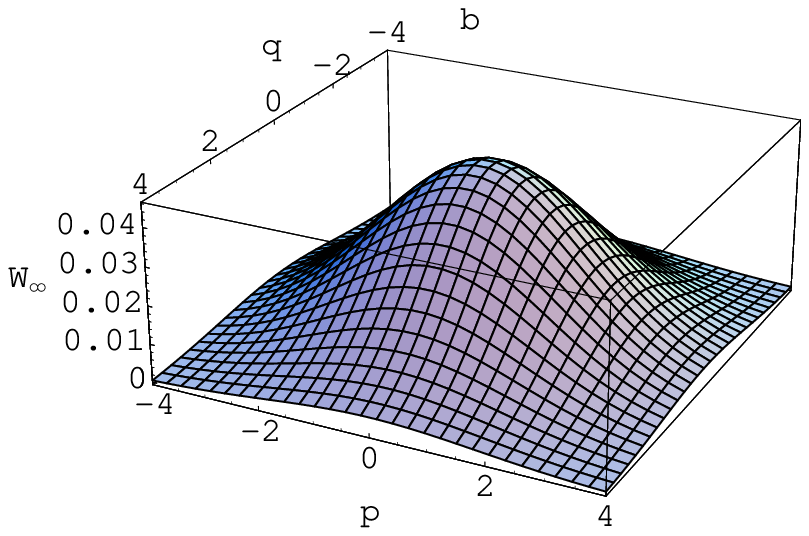,
width=0.6\textwidth}}\caption{Wigner function $W$ for
$\lambda=0.2,\mu=0.1,\delta=4,r=0$ and $C=3;$ a: $t=0;$
b: $t\to\infty.$}
\end{figure}

From expressions (\ref{qdec}) and (\ref{cor}) we notice
that the key parameter which describes QD and CC is
$\gamma.$ This coefficient determines the spread of the
Wigner function (\ref{wigc}) around the path in phase
space and measures the contribution of non-diagonal terms
in the density matrix (\ref{ccd3}). Therefore, when
decoherence increases, the correlations between the
canonical variables of coordinate and momentum decrease.
The extreme limit of QD ($\gamma\to\infty$) is
incompatible with CC and that of CC ($\gamma\to 0$) is
incompatible with QD. Their simultaneous realization is
not a trivial task: QD requires interaction with an
environment, which inevitably suppresses CC and produces
fluctuations in the evolution of the system, whereas
classical predictability requires these fluctuations to
be small. Therefore the existence of the environment is
crucial for the quantum to classical transition and,
consequently, classicality is an emergent property of an
open quantum system. The shown figures confirm the
presence of the relative competition which appears
between QD and existence of CC, since decoherence
(diagonalizing or the decreasing of the width of the
density matrix) implies a spreading of the Wigner
distribution function (which is the Fourier transform of
the density matrix) along the trajectory in phase space,
whereas CC require the existence of sharp peaks in the
Wigner function. Although there exists this competition,
there is a broad compromise regime in which QD and CC can
hold well simultaneously. If the density matrix is not
diagonal, but the Wigner function becomes peaked along
the classical trajectory for long times, we do not have,
strictly speaking, a classical limit, but only a
classical behaviour and CC. We regard classical behaviour
as a quantum behaviour in which there exists correlations
between coordinate and momentum or when coherences are
negligible, even without having CC.

We can assert that in the considered case classicality is a
temporary phenomenon, which takes place only at some stages of the
dynamical evolution, during a definite interval of time. Due to the
dissipative nature of evolution, the approximately deterministic
evolution is no more valid for very large times, when the
localization of the system is affected by the spreading of the wave
packet and of the Wigner distribution function.

In the case of a closed harmonic oscillator (zero-damping limit,
$\lambda=0$), $\delta_{QD}=1$ and the QD phenomenon does not take
place. At the same time we obtain (if $r=0$) \bea
\delta_{CC}(t)={2\over |(\delta-{1\over \delta})\sin(2\omega
t)|}.\eea Consequently, for an initial coherent state ($\delta=1$),
we see that $\delta_{CC}=\infty$ for any time, so that there are no
CC. If the initial state is squeezed ($\delta\neq 1$), then the
system manifests CC and it has a classical behaviour.

\section{Decoherence time scale and transition from quantum
mechanics to classical statistical mechanics}

Diffusion in momentum, which generates the decoherence in coordinate
$q,$ occurs at the rate set by $D_{pp} .$ In the macroscopic limit,
when $\hbar$ is small compared to other quantities with dimensions
of action, such as $\sqrt{D_{pp}<(q-q')^2>}$, the term in Eq.
(\ref{cooreq}) containing $D_{pp}/\hbar^2$ dominates and induces the
following evolution of the density matrix: \bea
{\partial\rho\over\partial t}=
-{D_{pp}\over\hbar^2}(q-q')^2\rho.\label{cooreq1}\eea Thus the
density matrix loses off-diagonal terms in position representation:
\bea<q|\rho(t)|q'>=<q|\rho(0)|q'>\exp[-{D_{pp}\over\hbar^2}(q-q')^2t],
\label{densol1} \eea while the diagonal ($q=q'$) ones remain
untouched. Quantum coherences decay exponentially at a rate given by
\bea {D_{pp}\over\hbar^2}(q-q')^2,\label{drate}\eea so that the
decoherence time scale is of the order of \bea
{\hbar^2\over{D_{pp}(q-q')^2}}.\eea In the case of a thermal bath,
we obtain (see Eq. (\ref{coegib})) \bea
t_{deco}={2\hbar\over{(\lambda+\mu)m\omega \sigma_{qq}(0)\coth
\epsilon}},\label{tdeco}\eea where we have taken $(q-q')^2$ of the
order of the initial dispersion in coordinate $\sigma_{qq}(0).$

In order to obtain a more precise expression of the decoherence
time, we consider the coefficient $\gamma$ (\ref{ccd4}), which
measures the contribution of non-diagonal terms in the density
matrix (\ref{ccd3}). For short times ($\lambda t\ll 1, \Omega t\ll
1$), we have: \bea \gamma(t)=-{m\omega\over
4\hbar\delta}\{1+2[\lambda(\delta+{r^2\over\delta(1-r^2)})\coth\epsilon
+\mu(\delta-{r^2\over\delta(1-r^2)})\coth\epsilon-\lambda-\mu-{\omega
r\over\delta\sqrt{1-r^2}}]t\}.\label{td}\eea From here we obtain
that quantum coherences in the density matrix decay exponentially at
a rate given by \bea
2[\lambda(\delta+{r^2\over\delta(1-r^2)})\coth\epsilon
+\mu(\delta-{r^2\over\delta(1-r^2)})\coth\epsilon-\lambda-\mu-{\omega
r\over\delta\sqrt{1-r^2}}]\eea and then the decoherence time scale
is \bea t_{deco}={1\over
2[\lambda(\delta+{r^2\over\delta(1-r^2)})\coth\epsilon
+\mu(\delta-{r^2\over\delta(1-r^2)})\coth\epsilon-\lambda-\mu-{\omega
r\over\delta\sqrt{1-r^2}}]}.\label{tdeco1}\eea The decoherence time
depends on the temperature $T$ and the coupling $\lambda$
(dissipation coefficient) between the system and environment
(through the diffusion coefficient $D_{pp}$), on the squeezing
parameter $\delta$ that measures the spread in the initial Gaussian
packet and on the initial correlation coefficient $r.$ We notice
that the decoherence time is decreasing with increasing dissipation,
temperature and squeezing.

For $r=0$ we obtain:\bea t_{deco}={1\over
2(\lambda+\mu)(\delta\coth\epsilon-1)}\label{tdeco2}\eea and at
temperature $T=0$ (then we have to take $\mu=0$), this becomes \bea
t_{deco}={1\over 2\lambda(\delta-1)}.\eea We see that when the
initial state is the usual coherent state $(\delta=1),$ then the
decoherence time tends to infinity. This corresponds to the fact
that for $T=0$ and $\delta=1$ the coefficient $\gamma$ is constant
in time, so that the decoherence process does not occur in this
case.

At high temperature, introducing the notation \bea \tau\equiv
{2kT\over \hbar\omega}\equiv {1\over\epsilon},\eea expression
(\ref{tdeco1}) becomes \bea t_{deco}={1\over
2[\lambda(\delta+{r^2\over\delta(1-r^2)})
+\mu(\delta-{r^2\over\delta(1-r^2)})]\tau}.\eea If, in addition
$r=0,$ then we obtain \bea t_{deco}={\hbar\omega\over
4(\lambda+\mu)\delta kT}.\eea

In Ref. \cite{unc} we studied the behaviour of the generalized
uncertainty function $\sigma(t)$ (\ref{sunc}). For short times we
obtained \bea \sigma(t)={\hbar^2\over 4}\{1+ 2[\lambda
(\delta+{1\over\delta(1-r^2)})\coth\epsilon+\mu(\delta-{1\over\delta(1-r^2)})
\coth\epsilon-2\lambda]t\}.\label{sunc1}\eea This expression shows
explicitly the contribution for small time of: (i) uncertainty that
is intrinsic to quantum mechanics, expressed through the Heisenberg
uncertainty principle and (ii) uncertainty due to the coupling to
the thermal environment, which has two components, dissipation and
diffusion (this last component is responsible for the process of
decoherence). From Eq. (\ref{sunc1}) we can determine the time $t_d$
when statistical (thermal) fluctuations become comparable with
quantum fluctuations. At high temperature we obtain \bea t_d={1\over
2\tau[\lambda
(\delta+{1\over\delta(1-r^2)})+\mu(\delta-{1\over\delta(1-r^2)})]}.\eea
By statistical (thermal) fluctuations we mean the fluctuations
produced by diffusion, that arise in the generalized uncertainty
function $\sigma(t)$ from the coupling of the harmonic oscillator to
the thermal bath at arbitrary temperature $T,$ even at $T=0$ (when
the diffusion coefficient still has a non-zero value). By quantum
fluctuations we mean fluctuations of the quantum harmonic oscillator
at zero coupling with the thermal bath.

As expected, we can see that the decoherence time
$t_{deco}$ has the same scale as the time $t_d$ after
which statistical fluctuations become comparable with
quantum fluctuations. The values of $t_{deco}$ and $t_d$
become closer with increasing temperature and squeezing.

When $t\gg t_{rel},$ where $t_{rel}\approx\lambda^{-1}$ is the
relaxation time, which governs the rate of energy dissipation, the
particle reaches equilibrium with the environment. Indeed, the
uncertainty function $\sigma(t)$ (\ref{sunc}) is insensitive to
$\lambda,\mu,\delta$ and $r$ and approaches \bea
\sigma^{BE}={\hbar^2\over 4}\coth^2\epsilon.\label{ube}\eea This is
the Bose-Einstein relation for a system of bosons in equilibrium at
temperature $T,$ obtained also in quantum Brownian models for the
weak coupling at arbitrary temperature. In the case of $T=0$ we
approach the limit of pure quantum fluctuations, \bea
\sigma_0={\hbar^2\over 4},\label{hut}\eea which is the quantum
Heisenberg relation. At high temperatures $T$ $(T\gg \hbar\omega/k)$
we obtain the limit of pure thermal fluctuations, \bea
\sigma^{MB}=({kT\over\omega})^2,\label{umb}\eea which is a
Maxwell-Boltzmann distribution for a system approaching a classical
limit.

For all macroscopic bodies the dissipation term becomes important
much later after the decoherence term has already dominated and
diminished the off-diagonal terms. If we compare the time scales of
these two terms (decoherence rate (\ref{drate}) and relaxation rate)
we get for high temperatures (for $\mu=0$) \bea {\rm
Decoherence~~rate \over Relaxation~~rate}={{D_{pp}(\Delta q)^2\over
\hbar^2}\over \lambda}= {m\omega\over
2\hbar}(q-q')^2\coth{\hbar\omega\over 2kT} \approx {mkT\over
\hbar^2}(q-q')^2.\eea In most typical situations this is a huge
number. For example, for a mass of 1 g at room temperature ($T=300$
K) and for a separation of $q-q'=1$ cm, the decoherence time scale
$t_{deco}$ is approximately 10$^{40}$ times shorter than the
relaxation time $t_{rel},$ so that in the macroscopic domain QD
occurs very much faster than relaxation. We remark also that
$t_{deco}$ can be of the order of $t_{rel}$ for sufficiently low
temperatures and small wave packet spread (small squeezing
coefficient). At the same time we have to remind that Lindblad
theory is obtained in the Markovian approximation, when the
characteristic time scales of the considered processes are larger
than the characteristic time scales of the thermal bath. When the
decoherence time scale is much shorter than the relaxation time and
at the same time the decoherence becomes faster even than the
environment time scales, then Markovian approximation is no more
valid and the use of non-Markovian quantum diffusion models becomes
preferable.

We have seen that a necessary condition for a system to behave
classically is the QD process. The time scale after which the system
has a classical behaviour, with an evolution described by a
classical probability distribution, is determined by the decoherence
time $t_{deco},$ when the density matrix becomes approximately
diagonal very rapidly during the system-environment interaction. On
the other hand, one often regards the regime where statistical
fluctuations begin to surpass quantum fluctuations as the transition
point from quantum to classical statistical mechanics and identifies
the high temperature regime of a system as the classical regime.
Above it was shown that these two criteria of classicality are
equivalent: the time when the quantum system decoheres is comparable
with the time when statistical fluctuations overtake quantum
fluctuations ($t_{deco}\approx t_{d}$). This result is a new
confirmation of the previous similar results \cite{AnH,hu1,hu2}.
However the regime after statistical fluctuations dominate should
not be called classical. After the decoherence time, although the
system is describable in terms of probabilities, it can not yet be
regarded as classical because of the spin-statistics effects. It has
to be described by non-equilibrium quantum statistical mechanics.
After the relaxation time the system is correctly treated by the
equilibrium quantum statistical mechanics, and only at a
sufficiently high temperature, when the spin (Fermi-Dirac or
Bose-Einstein) statistics can be represented by the
Maxwell-Boltzmann distribution function, it can be considered in a
classical regime \cite{unc,hu1,hu2} (see Eqs. (\ref{ube}),
(\ref{umb})).

\section{Summary and concluding remarks}

 We would like to remark that there are no precise quantitative
criteria in literature for classicality and at the same time there
exist some ambiguity and even unclarities concerning the
characteristics of the quantum to classical transition. In the
present paper we have studied QD and CC with the Markovian equation
of Lindblad in order to understand the transition from quantum to
classical mechanics for a system consisting of an one-dimensional
harmonic oscillator in interaction with a thermal bath in the
framework of the theory of open quantum systems based on quantum
dynamical semigroups. Our results may be summarized as follows.

(1) Using the criterion of QD for the considered model,
we have shown that QD in general increases with time and
temperature. For large temperatures, QD is strong and the
degree of mixedness is high, while for zero temperature
the asymptotic final state is pure. With increasing
squeezing parameter and initial correlation, QD becomes
stronger, but the asymptotic value of the degree of QD
does not depend on the initial squeezing and correlation,
it depends on temperature only.

(2) Using the criterion of CC, we have shown that the general
tendency is that CC are less and less strong with increasing time
and temperature. For a long enough interval of time we have a
significant degree of CC, but at $t\to\infty$ there are no CC in the
case of an asymptotic Gibbs state. With increasing squeezing
parameter, CC become stronger, but the variation of the correlation
coefficient has a small influence on the behaviour of the degree of
CC.

(3) During a finite interval of time the system interacting with the
thermal bath manifests simultaneously both QD and CC, so that a true
quantum to classical transition takes place and the system recovers
classicality in a significant measure. CC are expressed by the fact
that the Wigner function has a peak which follows (exactly for
$\lambda=\mu$ and approximately for $\lambda\neq\mu$) the classical
trajectory in phase space and QD is expressed by the loss of quantum
coherence in the case of a thermal bath at finite temperature. For
an initial Gaussian quantum state, Wigner function is positive for
all times, so that it represents a true classical probability
distribution in phase space.

(4) The expressions of the degree of QD and CC confirm
the relative competition which appears between QD and
existence of CC, since decoherence implies a spreading of
the Wigner distribution function along the trajectory in
phase space, whereas CC require the existence of sharp
peaks in the Wigner function. However, there exists a
broad compromise regime in which QD and CC can hold well
simultaneously. Consequently, classicality is a temporary
phenomenon, which takes place only at some stages of the
dynamical evolution of the system, during a definite
interval of time.

(5) We determined the general expression of the decoherence time,
which shows that it is decreasing with increasing dissipation,
temperature and squeezing. We have also shown that the decoherence
time has the same scale as the time after which statistical
fluctuations become comparable with quantum fluctuations, as
expected, and the values of these scales become closer with
increasing temperature and squeezing. After the decoherence time,
the decohered system is not necessarily in a classical regime. There
exists a quantum statistical regime in between. For the considered
open system at a finite temperature, the uncertainty relation
(\ref{ube}) holds, which interpolates between the Heisenberg
relation at zero temperature (\ref{hut}) and the high temperature
classical statistical relation (\ref{umb}). Only at a sufficiently
high temperature, when the spin statistics can be represented by the
Maxwell-Boltzmann distribution, the system can be considered in a
classical regime.

The study of classicality using QD and CC leads to a
deeper understanding of the quantum origins of the
classical world. As a result of the progress made in the
last two decades, the quantum to classical transition has
become a subject of experimental investigations, while
previously it was mostly a domain of theory
\cite{pa01,zu03}. The issue of quantum to classical
transition points to the necessity of a better
understanding of open quantum systems. The Lindblad
theory provides a self-consistent treatment of damping as
a general extension of quantum mechanics to open systems
and gives the possibility to extend the model of quantum
Brownian motion. The obtained results in the framework of
the Lindblad theory are a useful basis for the
description of the connection between uncertainty,
decoherence and correlations (entanglement) of open
quantum systems with their environment.

{\bf Acknowledgments}

Financial support and hospitality at the Institute of Theoretical
Physics in Giessen (Germany) during the stay of one (A. I.) of the
authors are gratefully acknowledged. The authors would also like to
thank the referee for useful criticism and recommendations.


\begin{thebibliography}{99}

\bibitem{gi96}
E. Joos, H. D. Zeh, C. Kiefer, D. Giulini, J. Kupsch and I. O.
Stamatescu, {\it Decoherence and the Appearance of a Classical World
in Quantum Theory,} 2nd Edn (Springer, Berlin, 2003)

\bibitem{pa01}
J. P. Paz and W. H. Zurek, in {\it Coherent Atomic Matter Waves, Les
Houches Session LXXII,} ed. by R. Kaiser, C. Westbrook and F. David
(Springer, Berlin, 2001), p. 533

\bibitem{zu03}
W. H. Zurek, Rev. Mod. Phys. {\bf 75}, 715 (2003)

\bibitem{mo90}
M. Morikawa, Phys. Rev. D {\bf 42}, 2929 (1990)

\bibitem{ha90}
S. Habib and R. Laflamme, Phys. Rev. D {\bf 42}, 4056 (1990)

\bibitem{ali}
R. Alicki, Open Sys. and Information Dyn. {\bf 11}, 53 (2004)

\bibitem{ze70}
H. D. Zeh, Found. Phys. {\bf 1}, 69 (1970); {\bf 3}, 109 (1973)

\bibitem{zu82}
W. H. Zurek, Phys. Rev. D {\bf 24}, 1516 (1981)
W. H. Zurek, Phys. Rev. D {\bf 26}, 1862 (1982)

\bibitem{jo84}
E. Joos, Phys. Rev. D {\bf 29}, 1626 (1984)

\bibitem{jo85}
E. Joos and H. D. Zeh, Z. Phys. B {\bf 59}, 223 (1985)

\bibitem{ca85}
A. O. Caldeira and A. J. Leggett, Phys. Rev. A {\bf 31}, 1059 (1985)

\bibitem{un89}
W. G. Unruh and W. H. Zurek, Phys. Rev. D {\bf 40}, 1071 (1989)

\bibitem{hu92}
B. L. Hu, J. P. Paz and Y. Zhang, Phys. Rev. D {\bf 45}, 2843 (1992)

\bibitem{pa93}
J. P. Paz, S. Habib and W. H. Zurek, Phys. Rev. D {\bf 47}, 488
(1993)

\bibitem{zu91}
W. H. Zurek, Phys. Today {\bf 44}, No. 10, 36 (1991); {\bf 46}, No.
12, 81 (1993); Prog. Theor. Phys. {\bf 89}, 281 (1993); {\it
Physical Origins of Time Asymmetry}, ed. by J. Halliwell, J.
Perez-Mercader and W. Zurek (Cambridge University Press, Cambridge,
1994)

\bibitem{br96}
M. Brune, E. Hagley, J. Dreyer, X. Ma\^ itre, A. Maali, C.
Wunderlich, J.-M. Raimond and S. Haroche, Phys. Rev. Lett. {\bf
77}, 4887 (1996)

\bibitem{ch99}
C. C. Cheng and M. G. Raymer, Phys. Rev. Lett. {\bf 82}, 4807 (1999)

\bibitem{am98}
H. Ammann, R. Gray, I. Shvarchuck and N. Christensen, Phys. Rev.
Lett. {\bf 80}, 4111 (1998)

\bibitem{kl98}
B. G. Klappauf, W. H. Oskay, D. A. Steck and M. G. Raizen, Phys.
Rev. Lett. {\bf 81}, 1203 (1998); Erratum in Phys. Rev. Lett. {\bf
82}, 241 (1999)

\bibitem{tu00}
Q. A. Turchette, C. J. Myatt, B. E. King, C. A. Sackett, D.
Kielpinski, W. H. Itano et. al, Phys. Rev. A {\bf 62}, 053807
(2000)

\bibitem{ko01}
D. A. Kokorowski, A. D. Cronin, T. D. Roberts, D. E. Pritchard,
Phys. Rev. Lett. {\bf 86}, 2191 (2001)

\bibitem{ni00}
M. A. Nielsen and I. L. Chuang, {\it Quantum Computation
and Quantum Information} (Cambridge Univ. Press,
Cambridge, 2000)

\bibitem{d}
E. B. Davies, {\it Quantum Theory of Open Systems} (Academic Press,
New York, 1976)

\bibitem{l1}
G. Lindblad, Commun. Math. Phys. {\bf 48}, 119 (1976)

\bibitem{s}
H. Spohn, Rev. Mod. Phys. {\bf 52}, 569 (1980)

\bibitem{l2}
G. Lindblad, Rep. Math. Phys. {\bf 10}, 393 (1976)

\bibitem{ss}
A. Sandulescu and H. Scutaru, Ann. Phys. (N.Y.) {\bf 173}, 277
(1987)

\bibitem{rev}
A. Isar, A. Sandulescu, H. Scutaru, E. Stefanescu and W. Scheid,
Int. J. Mod. Phys. E {\bf 3}, 635 (1994)

\bibitem{dodkur}
V. V. Dodonov, E. V. Kurmyshev and V. I. Man'ko, Phys. Lett. A {\bf
79}, 150 (1980)

\bibitem{unc}
A. Isar and W. Scheid, Phys. Rev. A {\bf 66}, 042117 (2002)

\bibitem{i2}
A. Isar, W. Scheid and A. Sandulescu, J. Math. Phys. {\bf 32}, 2128 (1991)

\bibitem{wig}
A. Isar, Helv. Phys. Acta {\bf 67}, 436 (1994)

\bibitem{vlas}
A. Isar, A. Sandulescu and W. Scheid, Int. J. Mod. Phys. B {\bf 10}, 2767
(1996)

\bibitem{for}
A. Isar, Fortschr. Phys. {\bf 47}, 855 (1999)

\bibitem{pur}
A. Isar, A. Sandulescu and W. Scheid, Phys. Rev. E {\bf 60}, 6371
(1999)

\bibitem{AnH}
C. Anastopoulos and J. J. Halliwell, Phys. Rev. D {\bf 51}, 6870
(1995)

\bibitem{hu1}
B. L. Hu and Y. Zhang, Mod. Phys. Lett. A {\bf 8}, 3575 (1993)

\bibitem{hu2}
B. L. Hu and Y. Zhang, Int. J. Mod. Phys. A {\bf 10}, 4537 (1995)

\end{thebibliography}
\end{document}